\documentclass[8pt]{elsart}

\usepackage{graphicx}
\usepackage{latexsym}
\usepackage{wrapfig}
\usepackage[rightcaption]{sidecap}
\usepackage{caption2}
\usepackage{amsmath}
\usepackage{mathrsfs}
\begin{document}
\begin{frontmatter}
\title{Quasinormal modes in
Schwarschild black holes due to arbitrary spin fields}
\author[shao,gscas]{Fu-Wen Shu}
\author[shao,nao,itp,email]{You-Gen Shen}
\address[shao]{Shanghai Astronomical Observatory, Chinese Academy of Sciences,
Shanghai 200030, People's Republic of China}
\address[nao]{National Astronomical Observatories, Chinese Academy of Sciences,
Beijing 100012, People's Republic of China}
\address[itp]{Institute of Theoretical Physics, Chinese Academy of Sciences,
Beijing 100080, People's Republic of China}
\address[gscas]{Graduate School of Chinese Academy of Sciences,
Beijing 100039, People's Republic of China.}
\thanks[email]{e-mail: ygshen@center. shao. ac. cn}

\begin{abstract}
The Newman-Penrose formalism is used to deal with the massless
scalar, neutrino, electromagnetic, gravitino and gravitational
quasinormal modes (QNMs) in Schwarzschild black holes in a united
form. The quasinormal mode frequencies evaluated by using the
3rd-order WKB potential approximation show that the boson
perturbations and the fermion
 perturbations behave in a contrary way for the
variation of the oscillation frequencies with spin, while this is
no longer true for the damping's,
 which variate with $s$ in a same way both for boson and fermion perturbations.
 \noindent PACS: number(s): 04.70.Dy, 04.70.Bw, 97.60.Lf 
\end{abstract}

\end{frontmatter}
%%%%%%%%%%%%%%%%%%%%%%%%%%%%%%%%%%%%%%%%%%%%%%%%%%%%%%%%%%%%%%%%%%%%%%
%%%%%%
%%%%%%%%%%%%%%%%
%%%%%%%%%%%%%%%%%%%%%%%%%%%%%%%%%%%%%%%%%%%%%%%%%%%%%%%%%%%%%%%%%%%%%%
%%%%%%
%%%%%%%%%%%%%%%%
\hspace*{7.5mm}Ever since Chandrasekhar\cite{CH1} and
 Vishveshwara\cite{VI} discovered the quasinormal modes of
 black holes, much effort has been devoted to investigating the
 QNMs of various black hole cases\cite{VC,CM,SS,EB,CHO}. By obtaining
 quasinormal mode(QN) frequencies, we can not only test the stability
 of the spacetime against small perturbations, but also probe
 the parameters of black hole, such as its mass, charge, and
 angular momentum, and thus, help uniquely identify a black
 hole.\\
 \hspace*{7.5mm}QNMs are described as the \textit{pure tones} of
 black hole. They are defined as solutions of the perturbation
 equations belonging to certain complex characteristic frequencies
 which satisfy the boundary conditions appropriate for purely ingoing
 waves at the event horizon and
purely outgoing waves at infinity\cite{CH2}. QNMs are excited by
the external perturbations(may be induced, for example, by the
falling matter). They appear as damped oscillations described by
the complex characteristic frequencies which are entirely fixed by
the properties of background geometery, and independent of the
initial perturbation. These frequencies can be detected by
observing the gravitational wave signal\cite{FE} : this makes QNMs
be of particular relevance in gravitational wave astronomy.
\\
\hspace*{7.5mm}QNMs were firstly used to study the stability of a
black hole. Detweiler and Leaver found the relations between the
parameters of the black hole and QNMs. Latest studies show that
QNMs play an important role in the quantum theory of gravity. This
is related to the quantization of black hole area\cite{JB}. For
example, there exit some possible relations between the classical
vibrations of black holes and various quantum aspects, such as the
relation between the real part of the quasinormal mode frequencies
and the Barbero-Immirzi parameter, a factor introduced by hand in
order that loop quantum gravity reproduces correctly entropy of
the black hole\cite{SH,OD,AC}. All these works deal with
asymptotically flat spacetimes. The recently proposed AdS/CFT
correspondence makes the QNMs more appealing, due to its argument
that string theory in anti-de Sitter(AdS) space is equivalent to
conformal field theory(CFT)in one less dimension\cite{JM}. In
addition to this context, many studies also have been done on QNMs
of various spin-fields\cite{VC,CHO,CH2} . Chandrasekhar\cite{CH2}
investigated the QNMs of fields with spin $s=1/2,1,2$ in Kerr
black hole, and $s=2$ in Schwarzschild and Reissner-Nordstr\"{o}m
black hole . Cardoso and Lemos\cite{VC} studied the QNMs of
Schwarzschild-anti-de Sitter for fields with spin $s=1,2$.
Cho\cite{CHO} calculated the massive Dirac quasinormal mode
frequencies of Schwarzschild black hole. However, problem on how
to deal with QNMs due to arbitrary spin field ($s=0,1/2,1,3/2,2$)
in a united form has never been discussed in these previous works
. The main purpose of this article is to give a possible way to
deal with QNMs
in Schwarzschild black hole of arbitrary spin field in a united form.\\
\hspace*{7.5mm}We start with the line element in standard
coordinates for the Schwarzschild space-time
\begin{equation}
ds^{2} =-e^{2U}dt^{2}+e^{-2U}dr^{2}+r^{2}\left( {d\theta ^{2} +
sin^{2}\theta d\varphi ^{2}} \right),
\end{equation}
with
\begin{equation}
 e^{2U}
 =1-\frac{{2M}}{{r}},\\
 \end{equation}
where $M$ is the mass of the black hole. \\
The Teukolsky's master equations\cite{SAT,SAT1} for massless
 arbitrary spin fields ($s=0,1/2,1,3/2,2$) in Newman-Penrose formalism can be written
 as\cite{GFT}
\begin{eqnarray}
\nonumber\{[D-(2s-1)\epsilon+\epsilon^{*}-2s\rho-\rho^{*}](\bar{\Delta}-2s\gamma+\mu)\\
\nonumber-[\delta-(2s-1)\beta-\alpha^{*}-2s\tau+\pi^{*}](\bar{\delta}-2s\alpha+\pi)\\
-(s-1)(2s-1)\Psi_{2}\}\Phi_{s}=0,\label{sp1}
 \end{eqnarray}
and
\begin{eqnarray}
\nonumber\{[\bar{\Delta}+(2s-1)\gamma-\gamma^{*}+2s\mu+\mu^{*}](D+2s\epsilon-\rho)\\
\nonumber-[\bar{\delta}+(2s-1)\alpha+\beta^{*}+2s\pi-\tau^{*}](\delta+2s\beta-\tau)\\
-(s-1)(2s-1)\Psi_{2}\}\Phi_{-s}=0.\label{sp2}
\end{eqnarray}
Assume that the wave-functions in Eqs. (\ref{sp1}) and (\ref{sp2})
have a $t$- and a $\varphi$-dependence specified in the form
$e^{i(\omega t+m\varphi)}$, i.e.,
\begin{equation}
 \Phi_{s}=R_{+s}(r)A_{+s}(\theta)e^{i(\omega t+m\varphi)},\quad \Phi_{-s}=R_{-s}(r)A_{-s}(\theta)e^{-i(\omega t+m\varphi)}.
 \end{equation}
and define
\begin{equation}
P_{+s}=\Delta^{s}R_{+s},\quad P_{-s}=r^{2s}R_{-s},
 \end{equation}
 (where $R_{\pm s}$ and $A_{\pm s}$ are, respectively, functions of
 $r$ and $\theta$ only, and $\Delta=r^2-2M$) one can decouple equations (\ref{sp1}) and (\ref{sp2}) as two pairs of equations,
\begin{eqnarray}
[\Delta\mathscr{D}_{1-s}\mathscr{D}^{\dag}_{0}-2(2s-1)i\omega
r]P_{+s}&=&\lambda P_{+s},\label{sp5}\\
\mathscr{L}^{\dag}_{1-s}\mathscr{L}_{s}A_{+s}&=&-\lambda
A_{+s},\label{sp6}
\end{eqnarray}
and
\begin{eqnarray}
[\Delta\mathscr{D}^{\dag}_{1-s}\mathscr{D}_{0}+2(2s-1)i\omega
r]P_{-s}&=&\lambda P_{-s},\label{sp7}\\
\mathscr{L}_{1-s}\mathscr{L}^{\dag}_{s}A_{-s}&=&-\lambda
A_{-s},\label{sp8}
\end{eqnarray}
where $\lambda$ is a separation constant. The reason we have not
distinguished the separation constants which derived from
Eqs.(\ref{sp1}) and (\ref{sp2}) is that $\lambda$ is a parameter
that is to be determined by the fact that $A_{+s}$ should be
regular at $\theta=0$ and $\theta=\pi$, and thus the operator
acting on $A_{-s}$ on the left-hand side of Eq.(\ref{sp8}) is the
same as the one on $A_{+s}$ in Eq.(\ref{sp6}) if we replace
$\theta$ by $\pi- \theta$. \\
In Schwarzschild black hole, the separation constant can be
determined analytically\cite{SAT1,NP}\\
for boson
\begin{equation}
\lambda=(l+|s|)(l-|s|+1),\quad\quad\quad l=|s|,|s|+1,\cdots
\end{equation}
for fermion
\begin{equation}
\lambda=(j+|s|)(j-|s|+1),\quad\quad\quad j=|s|,|s|+1,\cdots, and
\quad j=l\pm |s|
\end{equation}
where $l$ and $j$ represent angular quantum number and total
quantum number, respectively. Since $P_{+s}$ and $P_{-s}$ satisfy
complex-conjugate equations (\ref{sp5}) and (\ref{sp7}), it will
suffice to consider the equation
(\ref{sp5}) only.\\
By introducing a tortoise coordinate transformation
$dr_{*}=\frac{r^{2}}{\Delta}dr$, one can rewrite the operators
$\mathscr{D}_{0}$ and $\mathscr{D}^{\dag}_{0}$ as
\begin{equation}
\mathscr{D}_{0}=\frac{r^{2}}{\Delta}\Lambda_{+}, \quad and \quad
\mathscr{D}^{\dag}_{0}=\frac{r^{2}}{\Delta}\Lambda_{-}.
\end{equation}
where we have defined $\Lambda_{\pm}=\frac{d}{dr_{*}}\pm i\omega$. \\
With the definition $Y=r^{1-2s}P_{+s}$, equation (\ref{sp5}) can
be written as
\begin{equation}
\frac{r^{2s+3}}{\Delta}\left\{\Lambda^{2}Y+\left[\frac{d}{dr_{*}}\ln\frac{r^{4s}}
{\Delta^{s}}\right]\Lambda_{-}Y\right\}+\left[\Delta^{s}\frac{d}{dr}\left(\frac{1}{\Delta^{s-1}}
\frac{d}{dr}r^{2s-1}\right)-\lambda r^{2s-1}\right]Y=0,\label{sp9}
\end{equation}
where $\Lambda^2=\frac{d^2}{d(r_{*})^2}+\omega^2$. On further
simplification, Eq.(\ref{sp9}) can be brought to the form
\begin{equation}
\Lambda^{2}Y+P\Lambda_{-}Y-QY=0,\label{sp10}
\end{equation}
where
\begin{equation}
P=\frac{d}{dr_{*}}\ln\frac{r^{4s}}{\Delta^{s}},
\end{equation}
and
\begin{equation}
Q=\frac{\Delta}{r^{4}}\left[\lambda-(2s-1)(s-1)(\frac{2\Delta}{r^{2}}
-\frac{\Delta^{\prime}}{r})\right].\label{sp11}
\end{equation}
Considering the purpose of this article, we should seek to
transform Eq.(\ref{sp10}) to a one-dimensional wave-equation of
the form
\begin{equation}
\Lambda^{2}Z=VZ,\label{sp12}
\end{equation}
where $V$ represents potential. \\
\hspace*{7.5mm}The transformation theory introduced in
Ref.\cite{CH2} is applicable to solve this problem. We assume that
$Y$ is related to $Z$ in the manner
\begin{equation}
Y=\xi\Lambda_{+}\Lambda_{+}Z+W\Lambda_{+}Z,\label{sp13}
\end{equation}
where $\xi$ and $W$ are certain functions of $r_{*}$ to be
determined. One can then deduce the following equations\cite{CH2}
\begin{eqnarray}
\chi=\xi V+\frac{dT}{dr_{*}},\label{sp14}\\
\frac{d}{dr_{*}}\left(\frac{r^{4s}}{\Delta^{s}}\chi\right)=
\frac{r^{4s}}{\Delta^{s}}(QT-2i\omega\chi )+\beta,\label{sp15}\\
\chi\left(\chi-\frac{d
T}{dr_{*}}\right)+\frac{\Delta^{s}}{r^{4s}}\beta
T=\frac{\Delta^{s}}{r^{4s}}K,\label{sp16}\\
\chi V-Q\xi
V=\frac{\Delta^{s}}{r^{4s}}\frac{d\beta}{dr_{*}},\label{sp17}
\end{eqnarray}
where $K$ is a constant, and $\chi$, $T$
are certain functions of $r_*$.\\
\hspace*{7.5mm}The following work is to look for solutions of
equations (\ref{sp14})-(\ref{sp17}). These equations provide four
equations for five functions $\xi$, $\beta$, $\chi$, $T$, and $V$.
As a result, there is considerable difficulty in seeking useful
solutions of these equations. An obvious fact is that $\chi$ and
$V$ are independent of $\omega$ (i.e., they do not contain any
term linear in $i\omega$ ). Under these considerations, we can,
without loss of generality, suppose that $T$, $\beta$, $K$ are of
the forms\cite{CH2}
\begin{equation}
T=T_{1}(r_{*})+2i\omega f(s),
\quad\beta=\beta_{1}(r_{*})+2i\omega\beta_{2},\quad
K=\kappa_{1}+2i\omega\kappa_{2},\label{sp18}
\end{equation}
where $\beta_{2}$, $\kappa_{1}$, $\kappa_{2}$ are constants and
$f(s)$ is function of $s$. In this article, we take the choice
\begin{equation}
f(s)=\frac{1}{6}s(2s-1)(6s^{2}-23s+23)\quad\quad for\quad s=0,
\frac{1}{2}, 1, \frac{3}{2}, 2.
\end{equation}
Making use of the fact that, for a equation contains real and
imaginary parts, the real parts and imaginary parts in two sides
of the equation are equal, respectively, we can separate each of
the Eqs.(\ref{sp15})-(\ref{sp16}) into two equations by
substituting Eq.(\ref{sp18}), i.e.,
\begin{eqnarray}
 \chi=fQ+\frac{\Delta^{s}}{r^{4s}}\beta_{2},\label{sp19}\\
 \frac{d}{dr_{*}}\left(\frac{r^{4s}}{\Delta^{s}}\chi\right)=
 \frac{r^{4s}}{\Delta^{s}}QT_{1}+\beta_{1},\label{sp20}
 \end{eqnarray}
and
\begin{eqnarray}
\beta_{1}f+\beta_{2}T_{1}=\kappa_{2},\label{sp21}\\
\chi^{2}-\chi\frac{d
T}{dr_{*}}+\frac{\Delta^{s}}{r^{4s}}\beta_{1}T_{1}=\frac{\Delta^{s}}{r^{4s}}\kappa,\label{sp22}
\end{eqnarray}
where $\kappa=\kappa_{1}+4\omega^{2}f\beta_{2}$. \\
Substituting Eqs.(\ref{sp19}) and (\ref{sp21}) into
Eq.(\ref{sp20}), we obtain
\begin{equation}
T_{1}=\frac{f^{2}F_{,r_{*}}-\kappa_{2}}{fF-\beta_{2}}.
 \end{equation}
Here we have defined $F=\frac{r^{4s}Q}{\Delta^{s}}$, and `$,r_*$'
denotes the differential with respect to $r_{*}$. Eq.(\ref{sp22})
can then be written as
\begin{equation}
\frac{\Delta^{s}}{r^{4s}}(fF+\beta_{2})^{2}-f^{2}\frac{(fF+\beta_{2})
F_{,r_*,r_*}}{fF-\beta_{2}}+\frac{(f^{4}{F_{,r_*}}^{2}-\kappa_{2}^{2})F}
{(fF-\beta_{2})^{2}}=\kappa,\label{sp23}
 \end{equation}
It's obvious that Eq.(\ref{sp23}) is a condition on $F$ if
solutions of the chosen form are to exist, and hence the work to
seek available $\beta_{2}$, $\kappa$, and $\kappa_{2}$ whose
values satisfy Eq.(\ref{sp23}) is a key step to obtain the
function of potential $V$. Since $\kappa_{2}$ occurs as
$\kappa_{2}^{2}$ in Eq.(\ref{sp23}), two choices which associated
with $+\kappa_{2}$ and $-\kappa_{2}$ are possible to satisfy the
equation. Further study finds, an available choice is the
following
\begin{eqnarray}
\nonumber\beta_{2}=-\frac{1}{3}(2s-3)(s-2)(4s-1)\lambda,\\
\nonumber\kappa=\frac{1}{6}(2s-1)(s-1)(2s-3)(5s-8)\lambda(\lambda+s),\\
\kappa_{2}=
s(s-1)\left[\frac{8}{3}(2-s)\lambda\sqrt{s-\frac{1}{2}+\lambda}+(2s-1)(2s-3)M\right].
\end{eqnarray}
The solution for $V$ can then be obtained by substituting the
expressions of $\beta_{2}$, $\kappa$, and $\kappa_{2}$ into
Eq.(\ref{sp17}),i.e.,
\begin{equation}
V^{(\pm)}=\frac{\Delta^{s}}{r^{4s}}F-\frac{(fF-\beta_{2})fF_{,r_*,r_*}-f^{2}{F_{,r_*}}^{2}}
{(fF-\beta_{2})^{2}}\mp\frac{\kappa_{2}F_{,r_*}}{(fF-\beta_{2})^{2}},\label{potential}
\end{equation}
where we have distinguished the transformations associated with
$+\kappa_{2}$ and $-\kappa_{2}$ by superscripts $(\pm)$.\\
 \hspace*{7.5mm}It is obvious that $V^{(+)}$ equals to $V^{(-)}$ for $s=0,1$ from the equation (\ref{potential}).
 For the case of $s=\frac{1}{2},\frac{3}{2},2$, one can obtain
 a simpler form of $V^{(\pm)}$ by defining a new function
 $\tilde{F}$
\begin{equation}
V^{(\pm)}=\pm\kappa_{2}\frac{d\tilde{F}}{dr_{*}}+\kappa_{2}^{2}\tilde{F}^{2}+\kappa\tilde{F}.\label{sp24}
\end{equation}
The expression of this new function is
\begin{equation}
\tilde{F}=\frac{\Delta^{|s-1|}}{r^{4|s-1|}\left[\lambda+2(2s-1)(s-1)\frac{M}{r}\right]}.
\end{equation}
References\cite{CH2,And} has shown that potentials $V^{(+)}$ and
$V^{(-)}$ related in the way Eq.(\ref{sp24}) shows are equivalent
and hence possess the same spectra of quasinormal mode
frequencies. We shall therefore concentrate on $V^{(+)}$ only in
evaluating the quasinormal mode frequencies.\\
\hspace*{7.5mm}So we can simplify the radial equation (\ref{sp5})
to a one-dimensional wave-equation of the form
\begin{equation}
\frac{d^{2}Z}{dr_{*}^{2}}+\omega^{2}Z=VZ,
\end{equation}
with smooth real potentials, independent of $\omega$, i.e.,
\begin{equation}
V=\frac{\Delta^{s}}{r^{4s}}F-\frac{(fF-\beta_{2})fF_{,r_*,r_*}-f^{2}
{F_{,r_*}}^{2}+\kappa_{2}F_{,r_*}}{(fF-\beta_{2})^{2}},\label{sp25}
\end{equation}
where $F$ is a known function of $r_{*}$. Note that we have
written
 $V^{(+)}$ as $V$ because we shall not
 work with $V^{(-)}$, which will give the same quasinormal mode
 frequencies.\\
\hspace*{7.5mm}Figure.1 demonstrates the variation of the
effective potential $V$ with spin $s$. From this we can see that
peak values of the effective potential $V$ decrease with $s$ for
boson perturbations, while they increase with $s$ for fermion
perturbations. This phenomena is closely related to the value of
the separation constant $\lambda$.\\
\begin{SCfigure}[1][h]\centering
\includegraphics[width=2.7in,height=2.4in]{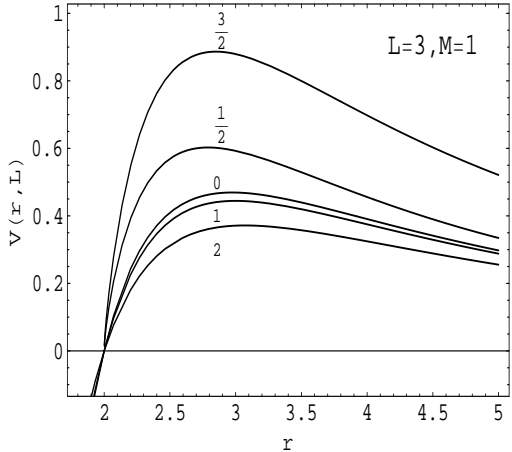}
\caption{Variation of the effective potential $V$ with spin $s$.}
\end{SCfigure}
\hspace*{7.5mm}Analytic expressions for the quasinormal mode
frequencies are usually very difficult to obtain. One hence should
appeal to approximation schemes to evaluate these frequencies.
Many methods are available for our purpose. One often used is WKB
approximation, a numerical method first proposed by
Mashhoon\cite{BM}, devised by Schutz and Will\cite{SW}, and was
subsequently extended to higher orders in\cite{SI1,RA}. An obvious
character of this scheme is that it is very accurate for low-lying
modes ($n<l$)\cite{SI2}. Under these considerations, we may use
WKB approximation to evaluate the quasinormal mode frequencies
within low-lying modes. The values
 for $n< l$ are listed in Tables 1 and 2, which also include a comparison with
 the results of Cho\cite{CHO}, and those of Chandrasekhar and Detweiler(CD)\cite{CH1}.
 Some gravitational modes ($l=3, n=2$, and $l=4, n=3$) of CD's results have no counterpart
in our results. This anomalous values were presumably
 results of numerical instabilities in their computational method that
 appeared when Im($\omega$)$\sim$ Re($\omega$) (see Ref.\cite{CH1} for details).
 From Table 1, we see that our results for boson
 perturbations are almost the same as Iyer's\cite{SI2}. Our values for Dirac
 modes agree well also with the results of Cho from Table 2.
 Notice that during our evaluating procedures,
  we have let the mass $M$ of the black hole
 as a unit of mass so as to simplify the calculation. For negative $n$, the values
 are related to these with positive $n$ by reflection of the
 imaginary axis.
 Figure 2 and figure 3 depict the variation of the QN frequencies with spin $s$. Some significant
 results can be found from these figures. (i) For boson perturbations, the real parts of the
 QN frequencies decrease with the spin value for the same $l$ and $n$,
 while they behave in a contrary way for the fermion
 perturbations. (ii) The imaginary parts of the frequencies
 decrease slowly with spin value, regardless of boson or fermion
 perturbations,
 as showed in Fig.3. This means that the variation of the
 oscillation frequencies with spin for boson perturbations is totally different from fermion
 perturbations, while the variation of the damping's with spin has
 a same law for both boson and fermion perturbations.\\
 \hspace*{7.5mm}The method we used to deal with QNMs of arbitrary spin
 fields in this article is also available for
 discussion on other black hole cases, such as Reissner-Nordstr\"{o}m and Kerr
 black hole, but a united form to deal with arbitrary spin
 perturbations in these black holes is still an unsolved problem.
\begin{center}\textbf{ACKNOWLEDGEMENTES}\end{center}
\hspace*{7.5mm}One of the authors(F.W. Shu) wishes to thank Doctor
Xian-Hui Ge for his valuable discussions. The work was supported
by the National Natural Science Foundation of China under Grant
No. 10273017.

\newpage
\begin{table}[t]\centering
\caption{Quasinormal mode frequencies for boson perturbations.
$\omega_i (i=0,1,2)$ represent scalar, electromagnetic and
gravitational perturbations, respectively. $\omega_{CD}$ represent
Chandrasekhar's results for gravitational perturbations.}
\begin{tabular}{cccccc}\hline\hline
$l$ & $n$ & $\omega_0$ & $\omega_1$& $\omega_2$ &$\omega_{CD}$\\
 \hline
 0 & 0 & 0.1046+0.1152i & & &\\
 1 & 0 & 0.2911+0.0980i &0.2459+0.0931i & &\\
 2 & 0 & 0.4832+0.0968i &0.4571+0.0951i & 0.3730+0.0891i & 0.3737+0.0890i\\
   & 1 & 0.4632+0.2958i &0.4358+0.2910i & 0.3452+0.2746i & 0.3484+0.2747i\\
 3 & 0 & 0.6752+0.0965i &0.6567+0.0956i & 0.5993+0.0927i & 0.5994+0.0927i\\
   & 1 & 0.6604+0.2923i &0.6415+0.2898i & 0.5824+0.2814i & 0.5820+0.2812i\\
   & 2 & 0.6348+0.4941i &0.6151+0.4901i & 0.5532+0.4767i & 0.4263+0.3727i\\
 4 & 0 & 0.8673+0.0964i &0.8530+0.0959i & 0.8091+0.0942i & 0.8092+0.0941i\\
   & 1 & 0.8557+0.2909i &0.8411+0.2893i & 0.7965+0.2844i & 0.7965+0.2844i\\
   & 2 & 0.8345+0.4895i &0.8196+0.4870i & 0.7736+0.4790i & 0.5061+0.4232i\\
   & 3 & 0.8064+0.6926i &0.7909+0.6892i & 0.7433+0.6783i &               \\
  \hline\hline
\end{tabular}
\end{table}
\begin{table}[t]\centering
\caption{Quasinormal mode frequencies for fermion
perturbations.$\omega_i (i=1/2,3/2)$ represent Dirac,
Rarita-Schwinger perturbations, respectively. $\omega_{Cho}$
represent Cho's results for Dirac perturbations. Notice that
$\kappa=l+1$ for $j=l+1/2$, according to Cho's definition.}
\begin{tabular}{cccccc}\hline\hline
$l$ & $\kappa$ & $n$ & $\omega_{1/2}$ & $\omega_{Cho}$&  $\omega_{3/2}$\\
 \hline
 1 &2& 0 & 0.3786+0.0965i & 0.379+0.097i&\\
 2 &3& 0 & 0.5737+0.0963i & 0.574+0.096i& 0.7346+0.0949i\\
   & & 1 & 0.5562+0.2930i & 0.556+0.293i& 0.7206+0.2870i\\
 3 &4& 0 & 0.7672+0.0963i & 0.767+0.096i& 0.9343+0.0954i\\
   & & 1 & 0.7540+0.2910i & 0.754+0.291i& 0.9233+0.2876i\\
   & & 2 & 0.7304+0.4909i & 0.730+0.491i& 0.9031+0.4835i\\
 4 &5& 0 & 0.9602+0.0963i & 0.960+0.096i& 1.1315+0.0956i\\
   & & 1 & 0.9496+0.2902i & 0.950+0.290i& 1.1224+0.2879i\\
   & & 2 & 0.9300+0.4876i & 0.930+0.488i& 1.1053+0.4828i\\
   & & 3 & 0.9036+0.6892i & 0.904+0.689i& 1.0817+0.6812i\\
  \hline\hline
\end{tabular}
\end{table}
\begin{figure}[h]\centering
\includegraphics[width=5in,height=3in]{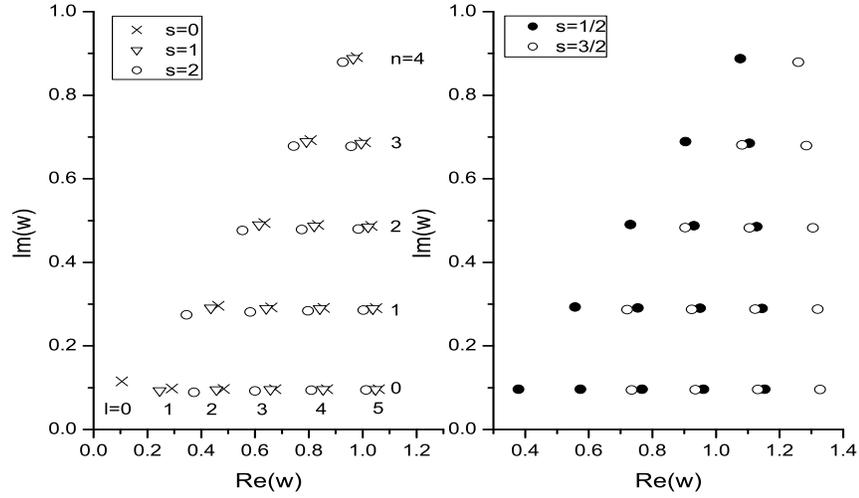}
\caption{Variation of the QN frequencies with spin $s$, the left
for boson perturbations, the right for fermion perturbations.}
\end{figure}
\begin{SCfigure}[1][h]\centering
\includegraphics[width=2.5in,height=2.5in]{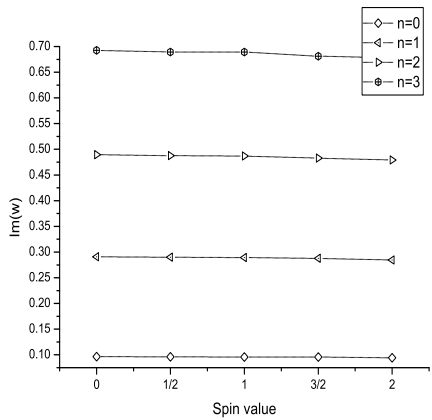}
\caption{Variation of the imaginary parts of the QN frequencies
with spin for $l=4$.}
\end{SCfigure}

\end{document}